\documentclass[11pt,twoside]{article}

\usepackage{asp2006}
\usepackage{lscape}
\usepackage{epsfig,graphicx,natbib,url}  

\begin{document}
\title{Space VLBI polarimetry of IDV sources: Lessons from VSOP and prospects for
VSOP2}   
\author{U.~Bach$^1$, T.P.~Krichbaum$^1$, S.~Bernhart$^{1,2}$,
C.M.V.~Impellizzeri$^1$, A.~Kraus$^1$, L.~Fuhrmann$^1$, A.~Witzel$^1$ \& J.A.~Zensus$^1$}   
\affil{$^1$Max-Planck-Institut fuer Radioastronomie, Bonn, Germany\\ $^2$IGG, University of Bonn, Germany}    

\begin{abstract} 

To locate and image the compact emission regions in quasars, which are closely
connected to the phenomenon of IntraDay Variability (IDV), space VLBI observations
are of prime importance. Here we report on VSOP observations of two prominent IDV
sources, the BL Lac objects S5\,0716+714. To monitor their short term variability,
these sources were observed with VSOP at 5\,GHz in several polarisation sensitive
experiments, separated in time by one day to six days, in autumn 2000.
Contemporaneous flux density measurements with the Effelsberg 100\,m radio
telescope were used to directly compare the single dish IDV with changes of the
VLBI images. A clear IDV behaviour in total intensity and linear polarization was
observed in 0716+714. Analysis of the VLBI data shows that the variations are
located inside the VLBI core component of 0716+714. In good agreement with the
single-dish measurements, the VLBI ground array images and the VSOP images, both
show a decrease in the total flux density of $\sim20$\,mJy and a drop of
$\sim5$\,mJy in the linear polarization of the VLBI core. No variability was found
in the jet. These findings are supported by VLBA observations of five IDV sources,
including 0716+714, in December 2000, that show a similar behaviour. From the
variability timescales we estimate a source size of a few micro-arcseconds and
brightness temperatures exceeding $10^{15}$\,K. Independent of whether the
interpretation of the IDV seen in the VLBI core is source intrinsic or extrinsic a
lower limit of $T_{\rm B} > 2\times10^{12}$\,K is obtained by model fitting of the 
VLBI-core. Our results show that future VSOP2 observations should be accompanied by
a single dish monitoring not only to discriminate between source-extrinsic
(interstellar scintillation) and source-intrinsic effects but to allow also a
proper calibration and interpretation of ultra-high resolution VSOP2 images.

\end{abstract}


\section{Introduction}   

Since the discovery of intraday variability (IDV, i.e. flux density and
polarization variations on time scales of less than 2 days) about 20 years ago
\citep{1986MitAG..65..239W,1987AJ.....94.1493H}, it has been shown that IDV is a
common phenomenon among extra-galactic compact flat-spectrum radio sources. It is
detected in a large fraction of this class of objects \citep[e.g.][see also
D.~Jauncey et al.; H.~Bignall et al., these
proceedings]{1992A&A...258..279Q,2001MNRAS.325.1411K,2003AJ....126.1699L}. The
occurrence of IDV appears to be correlated with the compactness of the VLBI source
structure on milliarcsecond scales: IDV is more common and more pronounced in
objects dominated by a compact VLBI core than in sources that show a prominent VLBI
jet. In parallel to the variability of the total flux density, variations in the
linearly polarized flux density and the polarization angle have been observed in
many sources 
\citep[e.g.][]{1989A&A...226L...1Q,1999NewAR..43..685K,2003A&A...401..161K,2004ChJAA...4...37Q}.
The common explanation for the IDV phenomenon at cm-wavelength is nowadays
interstellar scattering \citep[e.g.][]{1995A&A...293..479R,2001Ap&SS.278....5R}. On
the other hand some effects remain that cannot be easily explained by interstellar
scintillation and that are probably caused by relativistic jet physics
\citep[e.g.][]{1996ChA&A..20...15Q,2002ChJAA...2..325Q,2004ChJAA...4...37Q}. For
example the correlated intra-day variability between radio and optical wavelengths,
which is observed in sources like 0716+714 and 0954+658
\citep[e.g.][]{1990A&A...235L...1W,1991ApJ...372L..71Q,1996AJ....111.2187W} and the
recent detection of IDV at millimetre wavelengths in 0716+714 
\citep{2002PASA...19...14K,2003A&A...401..161K,2006A&A...456..117A,Fuhrmann2008}  suggests that
at least part of the observed IDV has a source-intrinsic origin.

Independent of the physical cause of IDV, it is obvious that IDV sources must
contain one or more ultra-compact emission regions. Using scintillation models,
typical source sizes of a few ten micro-arcseconds have been derived 
\citep[e.g.][]{1995A&A...293..479R,2002Natur.415...57D,2003ApJ...585..653B}. In the
case of source intrinsic variability and when using the light-travel-time argument,
even smaller source sizes of a few micro-arcseconds are obtained. In this case it
implies apparent brightness temperatures of up to $10^{18-19}$\,K (in exceptional
cases up to $10^{21}$\,K), far in excess of the inverse Compton limit of
$10^{12}$\,K \citep{1969ApJ...155L..71K}. These high apparent brightness
temperatures can be reduced e.g.\ by relativistic beaming with high Doppler-factors
\citep[e.g.][]{1991A&A...241...15Q,1996ChA&A..20...15Q,2002PASA...19...77K}.

The motivation of this VLBI monitoring, therefore, was to find how rapid structural
variability on sub-mas-scales occurs and where the IDV component is
located in the jet. An array of 12 antennas consisting of the 10 stations of NRAO's
VLBA, the 100\,m radio telescope of the Max-Planck-Institut f\"ur Radioastronomie
in Effelsberg (Germany), and the 8\,m HALCA antenna of the VSOP was used to follow
the short-term variability of 0716+714 and 0954+658 at 5\,GHz in autumn 2000. During short gaps in
the VLBI schedule the Effelsberg antenna was used to measure the light curve of a
calibrator and our target sources. A detailed description of the experiment and
data reduction can be found in \citet{2006A&A...452...83B}. Here we will
concentrate on the results of 0716+714, the results of 0954+658 will be presented
by Bernhart et al.~(in prep.).

\section{Results and Discussion}   

\begin{figure*}[htbp]
\hspace*{-5mm}
 \hbox{
 \includegraphics[bb=50 50 710 750,angle=0,width=4.7cm,clip] {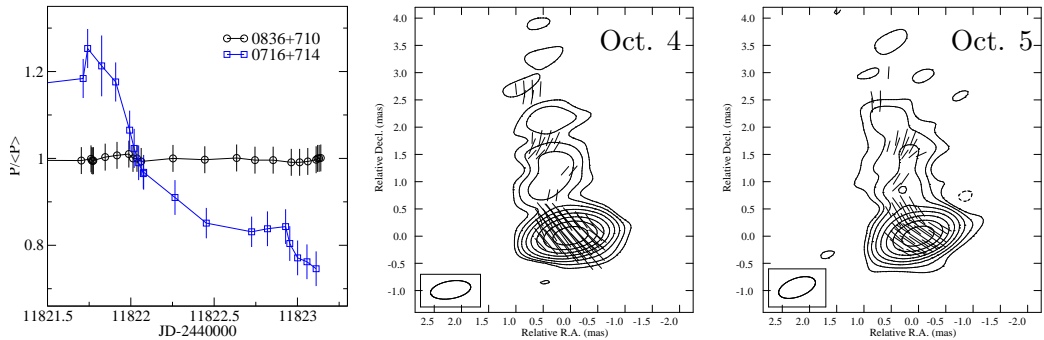}
 \includegraphics[angle=0,width=4.5cm] {bach_fg2.ps}
 \includegraphics[angle=0,width=4.5cm] {bach_fg3.ps}
 \put(-155,115){\makebox(0,0){Oct. 4}}
 \put(-22,115){\makebox(0,0){Oct. 5}}}
   \caption{{\bf Left:} Single dish polarization variability of 0716+41 on Oct. 4
and 5. {\bf Middle and right:} VSOP contour maps of
   Stokes $I$ of 0716+714 with polarization vectors superimposed (1\,mas
   corresponds to 6.7\,mJy/beam). Contours start at 1.8\,mJy/beam
   increasing in steps of 2. A clear drop in polarized flux (shorter vectors) and a
rotation by 10$^\circ$ is visible in the core region from Oct.~4 to 5.}\label{bach_fig:images}
\end{figure*}

The ground array and the VSOP images of 0716+714 show a bright core and a jet
oriented to the north (Fig.~\ref{bach_fig:images}). The linear polarization images
indicate that the jet magnetic field is perpendicular to the jet axis. Compared to
the jet axis, the electric vector position angle in the core is misaligned by
around 60\,$^\circ$. This is explained either by opacity effects in the core region
or by a curved jet. Here, jet curvature is supported by recent high resolution
3\,mm VLBI observations that show the inner jet structure ($r<0.1$\,mas) at a
similar position angle as the EVPA in the core at 6\,cm wavelength
\citep{2006A&A...452...83B}.

Simultaneous flux-density measurements with the 100\,m Effelsberg telescope during
the VSOP observations revealed variability in total intensity ($\sim 5$\,\%) and in
linear polarization (up to $\sim 40$\,\%) accompanied by a rotation of the
polarization angle by up to  $15\,^\circ$. The analysis of the VLBI data shows
that the intra-day variability in 0716+714 is associated with the VLBI-core region
and not with the milli-arcsecond jet. Both the ground array and the VSOP maps show a
similar decrease of the flux densities in total intensity and linear polarization
of the core component, which is in good agreement with the variations seen with the
Effelsberg radio-telescope \citep[more details are given in][]{2006A&A...452...83B}. These findings could be confirmed by VLBA observations
of five IDV sources, including 0716+714, a few month later in December 2000, that
show a similar behaviour (Impellizzeri et al.\ in perp.). In this, 0716+714
displays a behaviour that is similar to what was previously observed in the IDV
sources 0917+624 and 0954+658, where components in or near the VLBI core region
were also made responsible for the IDV \citep{2000MNRAS.315..229G}. Over the time
interval of our VSOP observations, no rapid variability in the jet was observed and
we cannot confirm the variability outside the core and in the jet found by
\cite{2000MNRAS.313..627G}.

The simultaneous variation of the polarization angle with the polarized intensity
in the core suggests that the variations might be the result of the sum of the
polarization of more than one compact sub-component on scales smaller than the beam
size (multi-comp.\ model). Assuming a redshift of $z = 0.3$ \citep{1996AJ....111.2187W} and that these
variations are intrinsic to the source, we derived brightness temperatures of $\sim
3\times 10^{15}$\,K to $\sim 10^{16}$\,K. Doppler factors of $>20$ are needed to
bring these values down to the inverse-Compton limit. These numbers agree with the
observed speeds in the jet if the angle to the line of sight is very small
($\theta<2^\circ$), as already proposed by \citet{2005A&A...433..815B}. Because of
the unknown redshift, the derived speeds and brightness temperatures represent only
lower limits.

However, interstellar scintillation effects could also explain the IDV seen in the 
VLBI core, if the core region consists of several compact and polarized
sub-components, with sizes of a few ten micro-arcseconds. To explain the observed
polarization variations, the sub-components must scintillate independently in a
different manner, which means that they  must have slightly different intrinsic
sizes and intrinsic polarization
\citep[e.g.][]{2001Ap&SS.278....5R,2001Ap&SS.278..129R}.

Independent of whether the interpretation of the IDV seen in the VLBI core is
source intrinsic or extrinsic, the space-VLBI limit to the core size ($<0.1$\,mas)
gives a robust lower limit to the brightness temperature of $\geq  2\times
10^{12}$\,K and therewith exceeds the inverse-Compton limit. This implies a lower
limit to the Doppler factor of about $\geq 4$ and, independent of the model we use
to explain the variability, relativistic beaming must play a role.

The increased sensitivity, higher resolution, and frequency flexibility of VSOP-2
will provide a powerful tool to look even deeper into the core region and to
distinguish which fraction of the IDV is due to source intrinsic variations and
which is caused by the interstellar medium.  Variability surveys like MASIV
\citep{2003AJ....126.1699L} revealed that IDV is present in a huge number of
sources and therefore it seems advisable that future VSOP2 experiments are
accompanied by a simultaneous flux monitoring to allow a proper calibration of the
data and to interpret possible structural changes on short time scales.




\end{document}